\documentclass{article}
\usepackage{spconf,amsmath,graphicx}

\usepackage{mathtools}

\DeclarePairedDelimiter\ceil{\lceil}{\rceil}

\usepackage{url}
\usepackage{listings}
\usepackage{caption}
\usepackage{tabularx}
\usepackage{tikz}
\usetikzlibrary{arrows.meta}

\usepackage{amsmath,bm, amssymb, setspace, booktabs}
\usepackage{multirow,bigdelim}

\usepackage{makecell}
\usepackage{xcolor,colortbl}
\usepackage{color}
\usepackage{colortbl}
\usepackage{hyperref}
\usepackage[shortlabels]{enumitem}

\usepackage[]{footmisc}
\newcommand{\zhenk}[1]{{\color{black}{#1}}}
\definecolor{creamson}{RGB}{153, 0, 0}

\usepackage{graphicx}

\usepackage{subfigure}
\input{def.set}

\begin{document}

\title{
sub-8-bit quantization for on-device speech recognition: \\a regularization-free approach
}


\name{Kai Zhen, Martin Radfar, Hieu Nguyen, Grant P. Strimel, Nathan Susanj, Athanasios Mouchtaris}
\address{Amazon Alexa AI}



\maketitle

\begin{abstract}
For on-device automatic speech recognition (ASR), quantization aware  training (QAT) is ubiquitous to achieve the trade-off between model predictive performance and efficiency. Among existing QAT methods, one major drawback is that the quantization centroids have to be predetermined and fixed. To overcome this limitation, we introduce a regularization-free, ``soft-to-hard" compression mechanism with self-adjustable centroids in a $\mu$-Law constrained space, resulting in a simpler yet more versatile quantization scheme, called General Quantizer (GQ). We apply GQ to ASR tasks using Recurrent Neural Network Transducer (RNN-T) and Conformer architectures on both LibriSpeech and de-identified far-field datasets. Without accuracy degradation, GQ can compress both RNN-T and Conformer into sub-8-bit, and for some RNN-T layers, to 1-bit for fast and accurate inference. We observe a $30.73\%$ memory footprint saving and $31.75\%$ user-perceived latency reduction compared to 8-bit QAT via physical device benchmarking. 
\end{abstract}

\begin{keywords}
On-device speech recognition, quantization aware  training, RNN-T, conformer, model efficiency
\end{keywords}

\section{Introduction}
Improving the efficiency of neural automatic speech recognition (ASR) models via quantization is critical for on-device deployment scenarios \cite{nguyen2020quantization,swaminathan2021codert}. 
For neural network accelerator (NNA) embedded devices, where memory and bandwidth are at a premium, quantization can reduce the footprint and lower the bandwidth consumption of ASR execution, which will not only afford a faster model inference but also facilitate model deployment to various portable devices where a stable network connection is limited.

Existing quantization methods can be post-training quantization (PTQ) or in-training / quantization aware training (QAT). PTQ is applied after the model training is complete by compressing models into 8-bit representations and is relatively well supported by various libraries \cite{li2021brecq,liu2021post,nahshan2021loss,wang2020towards, fasoli20214, kim2022integer}, such as TensorFlow Lite \cite{abadi2016tensorflow} and AIMET \cite{siddegowda2022neural} for on-device deployment. However, almost no existing PTQ supports customized quantization configurations to compress machine learning (ML) layers and kernels into sub-8-bit (S8B) regimes \cite{ding20224}. Moreover, the performance drop is inevitable as the model is unaware of the loss of precision when being quantized at test time. In contrast, QAT performs bit-depth reduction of model weights (for example, from 32-bit floating point to 8-bit integer) during training which usually yields superior performance over PTQ \cite{xu2021mixed}\cite{zafrir2019q8bert}.
The QAT mechanism can be in the forward pass (FP-QAT) or the backward pass (BP-QAT), with the difference being whether regularization is used in the loss function. FP-QAT \cite{ding20224} quantizes the model weights during forward propagation to pre-defined quantization centroids. BP-QAT \cite{nguyen2020quantization, zhen2022sub, strom2022squashed} relies on customized regularizers to gradually force weights to those quantization centroids (i.e., ``soft quantization" via gradient) during training before hard compression performs in the late training phase. 
As model weights are informed by the customized regularizers to move closer to where they are quantized at runtime per training step, the predictive performance is often well preserved.
Therefore, the focus of this work is on QAT.  

Under both FP- and BP-QAT, it is essential that the quantization centroids are defined and specified before model training. As such, the demerit is the low feasibility when quantizing models in 
S8B
mode because one needs to select the proper quantization centroids  and their configurations for each kernel in each layer to ensure minimal runtime performance degradation. 
Consequently, applying existing QAT methods to Conformer \cite{gulati2020conformer} becomes quite challenging, as it usually contains more than hundreds of kernels.

In this work, we propose General Quantizer (GQ), a regularization-free, model-agnostic quantization scheme with a mixed flavor of both FP- and BP-QAT. GQ is ``general" in that it does not augment the objective function by introducing any regularizer as in BP-QAT but determines the appropriate quantization centroids during model training for a given bit depth, and it can be simply applied in \zhenk{a plug-and-play manner to an arbitrary ASR model}. Unlike FP-QAT, GQ features a soft-to-hard quantization during training, allowing model weights to hop around adjacent \zhenk{partitions} more easily. \zhenk{Under GQ, quantization centroids are self-adjustable but in a $\mu$-Law constrained space.}
As a proof-of-concept, we adopt the ASR task and conduct \zhenk{experiments} on both \zhenk{the} LibriSpeech and de-identified far-field datasets to evaluate GQ on three major end-to-end ASR architectures, namely conventional Recurrent Neural Network Transducer (RNN-T) \cite{rao2017exploring}, Bifocal RNN-T \cite{macoskey2021bifocal}, and \zhenk{Conformer} \cite{li2021better}\cite{sainath2022improving}. Our results show that in all three architectures, GQ yields little to no accuracy loss when compressing models to 
S8B
or even sub-5-bit (5-bit or lower). We also present performance optimization strategies from ablation studies on bit-allocation and quantization frequency. Our contributions are as follows:
\begin{itemize}
    \item We propose GQ, inspired by both FP- and BP-QAT approaches. GQ enables on-centroid weight aggregation without augmented regularizers. Instead, it leverages Softmax annealing to impose soft-to-hard quantization on centroids from the $\mu$-Law constrained space. 
    \item GQ supports different quantization modes for a wide range of granularity: different bit depths can be specified for different kernels/layers/modules.
    \item With GQ, we losslessly compress a lightweight streaming Conformer into sub-5-bit with more than $6\times$ model size reduction. To our best knowledge, this is among the first sub-5-bit Conformer models for on-device ASR. Without accuracy degradation, our GQ-compressed 5-bit Bifocal RNN-T reduces the memory footprint by $30.73\%$ and P90 user-perceived latency (UPL) by $31.30\%$.
\end{itemize}

We describe the problem in Sec. \ref{sec:prob} and GQ in Sec. \ref{sec:algo}. The experimental settings and results are detailed in Sec. \ref{sec:exp}. We conclude in Sec. \ref{sec:conclude} with some final remarks. 

\section{Preliminaries}
\label{sec:prob}
\subsection{Problem Formulation}
Consider a general deep neural network architecture with $K$ layers, $\mathcal{F}=\mathcal{F}_1\circ\dots\circ\mathcal{F}_K$, mapping the input from $\mathbb{R}^{d_1}$ to the output in $\mathbb{R}^{d_{K+1}}$ as $\mathcal{F}:\mathbb{R}^{d_1}\longmapsto\mathbb{R}^{d_{K+1}}$, where the input and output of an arbitrary \zhenk{$k$-th} layer are \zhenk{$\boldsymbol{x}^{(k+1)} := \mathcal{F}_k(\boldsymbol{x}^{(k)})$}. Under supervised learning, the training data $\mathcal{X}=\{\boldsymbol{x_1}, \dots, \boldsymbol{x}_N\}$ and $\mathcal{Y}=\{\boldsymbol{y_1}, \dots, \boldsymbol{y}_N\}$ are \zhenk{used} for updating model weights $\mathbb{W}=[\boldsymbol{W}_1, \dots, \boldsymbol{W}_K]$ for $K$ layers in $\mathcal{F}$.
Usually the optimization process is over the training objective \zhenk{function} $\mathcal{L}(\mathcal{X},\mathcal{Y},\mathcal{F},\mathbb{W})=\frac{1}{N}\sum\limits_{i=1}^N\ell(\mathcal{F}(\boldsymbol{x_i}),\boldsymbol{y_i})+\lambda R(\mathbb{W})$, where \zhenk{$i$ is the data batch index}, $\ell$ is the major loss term measuring model accuracy and $R(\mathbb{W})$ is the regularizer blended to the objective function via a coefficient $\lambda$. 


Network quantization aims at discretizing model weights. For scalar quantization, it is to convert each weight,   $w\in\mathbb{W}$,  to a quantization centroid,  $z\in\boldsymbol{z}$, where $\boldsymbol{z} = [z_1, ... z_m]$ to ensure the network is compressed into $\ceil*{log_2 m}$-bit. For 
S8B
quantization, centroids are from a subset of \texttt{INT8} values.

\subsection{Related QAT Approaches}
BP-QAT counters model weight continuity via regularization. For example, it introduces weight regularizers on model weights, i.e., $R(\mathbb{W})$, measuring the point-wise distance between each weight and $m$ quantization centroids in the centroid vector $\boldsymbol{z} = [z_1, ... z_m]$.
Note that the quantization weight regularizer in the loss function\zhenk{, as $R(\mathbb{W})$,} must be gradient descent compatible.
Consequently, $R(\mathbb{W})$ cannot enforce each weight to be replaced by the closest centroid in $\boldsymbol{z}$ as $w = \arg\min\limits_i ||w-z_i||$ , for $w \in \mathbb{W}$ and $z_i \in \boldsymbol{z}$, because the $\min$ operator is not differentiable.
Recent BP-QAT methods force weights to approach the centroid in $\boldsymbol{z}$ using $R(\mathbb{W}) = \sum\limits_{w \in \mathbb{W}} \mathcal{D}(w,\boldsymbol{z}),$
where the differentiable dissimilarity function $\mathcal{D}$ is based on a cosine function in \cite{nguyen2020quantization,zhen2022sub}.

In contrast, FP-QAT can be regularizer free \cite{ding20224,lou2019autoq}. Usually, the process is to use a ``fake quantizer''  or equivalent operations during training, hard quantizing weights to a specific range and bit-depth; and then at runtime, converting the model to \texttt{INT8} format via TFLite \cite{david2021tensorflow}. The study \cite{ding20224} uses native quantization operators with which, during training, the weights are quantized and then converted to the integer type for model deployment. However, FP-QAT is essentially hard compression recurring during training with severely dropped performance when applied to 
S8B
quantization. Consequently, finetuning is usually needed, which prolongs the model training time \cite{martinez2021permute, piao2022sensimix}.

Both FP-QAT and BP-QAT require specifying appropriate quantization centroids before model training. While the centroids for \texttt{INT8} model compression are pre-defined, for 
S8B
quantization, the optimal set of centroids is usually kernel/layer specific. For models, such as Conformer \cite{gulati2020conformer},  where there are usually hundreds of kernels, current 
S8B
QAT methods become less tractable.

In this work, we combine the merit from both FP- and BP-QAT and propose General Quantizer (GQ) that navigates weights to quantization centroids without introducing augmented regularizers but via feedforward-only operators. 
Our work is inspired by a continuous relaxation of quantization \cite{agustsson2017soft} also used for speech representation learning \cite{zeghidour2021soundstream,ZhenK2019interspeech,jiang2022cross, petermann2021harp, cheon2021coded, zhen2021scalable, zhen2020efficient}, and $\mu$-Law algorithm for 8-bit pulse-code modulation (PCM) digital telecommunication \cite{kaneko1970unified}.

\section{Methods}
\label{sec:algo}
\subsection{Centroid Selection via Softmax-Based Dissimilarity Matrices}
\begin{figure}[t]
	\centering
	\subfigure[Weight value before and after quantization approximation. Here, the larger the softmax temperature $\alpha$ is, the closer to true quantization that the weight transformation becomes.]{\includegraphics[height=1.96in]{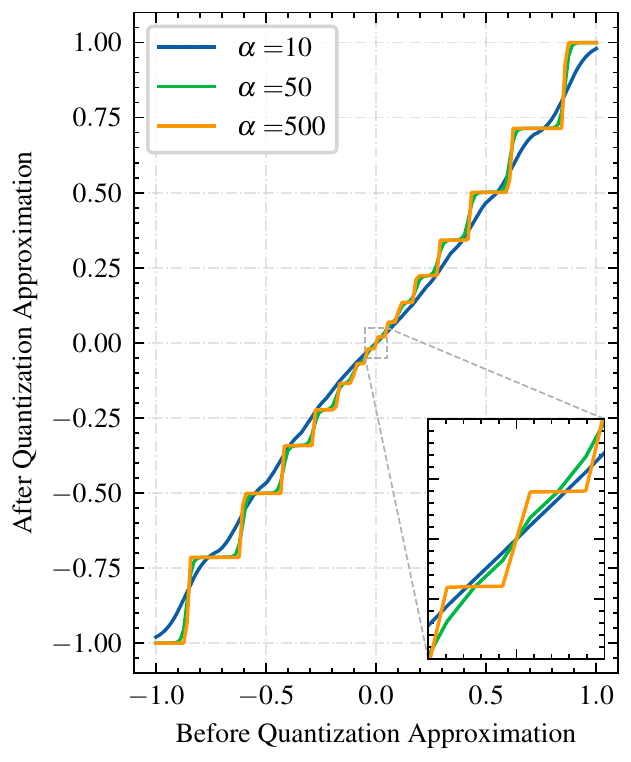}} 
	\hfill
  \subfigure[Relationship between the wrapped quantization centroids and $\mu$. A large $\mu$ means more quantization centroids are allocated near 0 and vice versa.]{\includegraphics[height=1.96in]{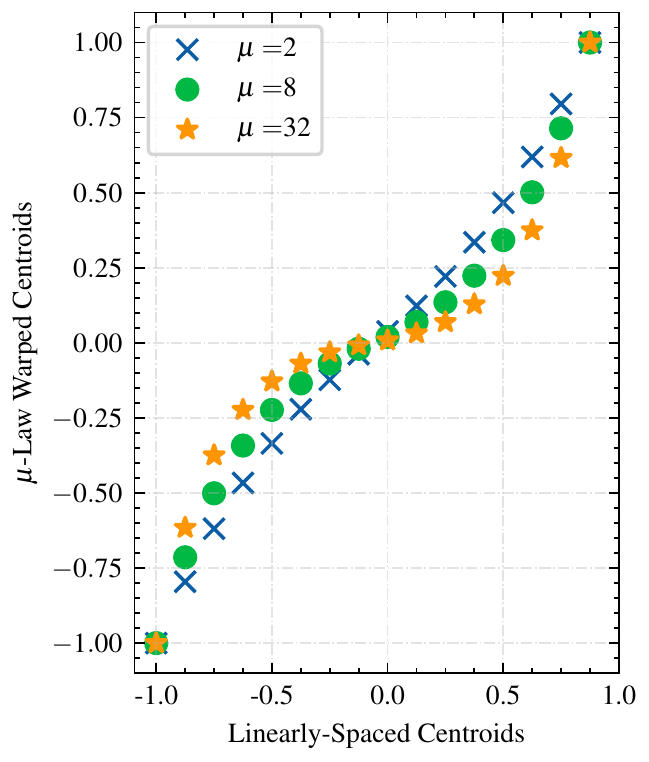}} 
	\caption{Hyperparameters in GQ: $\alpha$ adjusts the quantization temperature which increases gradually during training, and $\mu$ modulates the non-linearity of quantization centroids.}
	\label{fig:s2h}
	\vspace{-0.12in}
\end{figure}

For any weight value $w_i\in\boldsymbol{w}$ where $|\boldsymbol{w}|=n$, and the quantization centroid vector $\boldsymbol{z} = [z_1, ... z_m]$, we define the point-wise dissimilarity matrix in Eq. \ref{eq:dist}
\begin{equation}
\label{eq:dist}
    \boldsymbol{A}_{\text{soft}} =\begin{bmatrix} 
    a_{11} & \cdots & a_{1m}\\
    \vdots & \ddots  &\vdots\\
    a_{n1} & \cdots & a_{nm}
    \end{bmatrix},
\end{equation}
where $a_{ij}$ is the probability of representing $w_i$ by $z_j$. Each row in $\boldsymbol{A}_{\text{soft}}[i\cdot]$ is summed to $1$ with the largest probability going to the closest centroid. This is achieved when the point-wise distance $||w_i-z_j||_1$ is scaled by a negative number $-\alpha$ and wrapped by a Softmax function in Eq. \ref{eq:softmax} 
\begin{equation}
\label{eq:softmax}
    a_{ij} = \frac{e^{-\alpha||w_i-z_j||_1}}{\sum\limits_{j=1}^{m}e^{-\alpha||w_i-z_j||_1}}.
\end{equation}
Here, $\alpha \in [1, \infty)$ serves as the Softmax temperature for quantization annealing. When $\alpha$ is relatively small, $w_i$ will be approximated by all centroids in $\boldsymbol{z}$ (see Eq.\ref{eq:linear-approx}); when $\alpha\rightarrow\infty$, $\boldsymbol{A}_{\text{soft}}[i\cdot]$ becomes a one-hot vector $\boldsymbol{A}_{\text{hard}}[i\cdot]$ and the weight will be the closest centroid.
\begin{equation}
\label{eq:linear-approx}
    \overline{w_{i}} = \boldsymbol{z}\times{\boldsymbol{A}_{\text{soft}}[i\cdot]}^{T}.
\end{equation}
For simplicity, during training, we set the initial and target scalar values to be $\alpha_{\text{start}}$ and $\alpha_{\text{end}}$, and allow $\alpha$ to gradually and linearly increase from $s_{\text{start}}$ to $s_{\text{end}}$, as shown in Eq. \ref{eq:annealing}.
\begin{equation}
\label{eq:annealing}
    \alpha = \alpha_{\text{start}} + (s-s_{\text{start}})\times\frac{\alpha_{\text{end}}-\alpha_{\text{start}}}{s_{\text{end}}-s_{\text{start}}}.
\end{equation}
As a result, the QAT effect is gradually intensified. At $\alpha=10$, a rather small value, weights after being approximated by quantization centroids in $\boldsymbol{z}$ roughly preserve their original values; however, 
as $\alpha$ gradually increased to 500, the near-linear line almost becomes a step function, aggregating weights to just a few centroids (see Fig. \ref{fig:s2h} (a)).
This forms a soft-to-hard QAT and allows model weights to be updated via gradients with barely any extra constraint during the early stage of training before driving weights to a certain centroid.

\begin{figure}[h]
	\centering
	{\includegraphics[width=0.95\columnwidth]{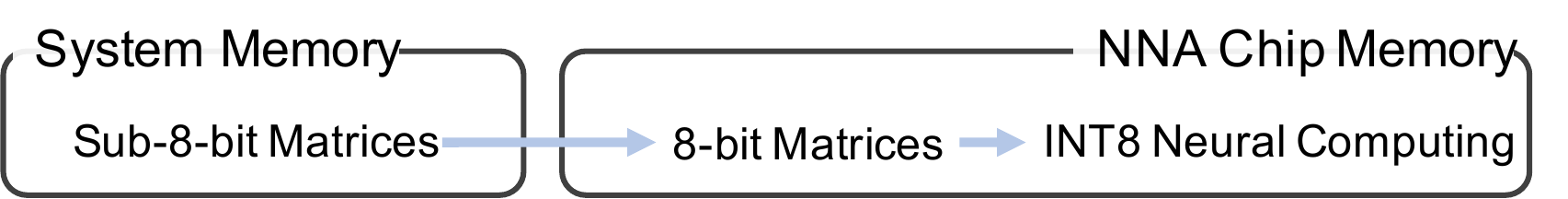}}	
	\caption{Loading weights from system memory to chip memory on NNA is faster in 
S8B
format, although the arithmetic operation is still in \texttt{INT8} format.}
	\label{fig:nna}
	\vspace{-0.12in}
\end{figure}

\begin{figure*}[t]
	\centering
\subfigure[During model training, the approximation scalar $\alpha$ increases, gradually and more effectively aggregating adjacent weights toward the corresponding centroid. $\mu$ is adaptable as model weights converge.]{\includegraphics[width=0.57\linewidth]{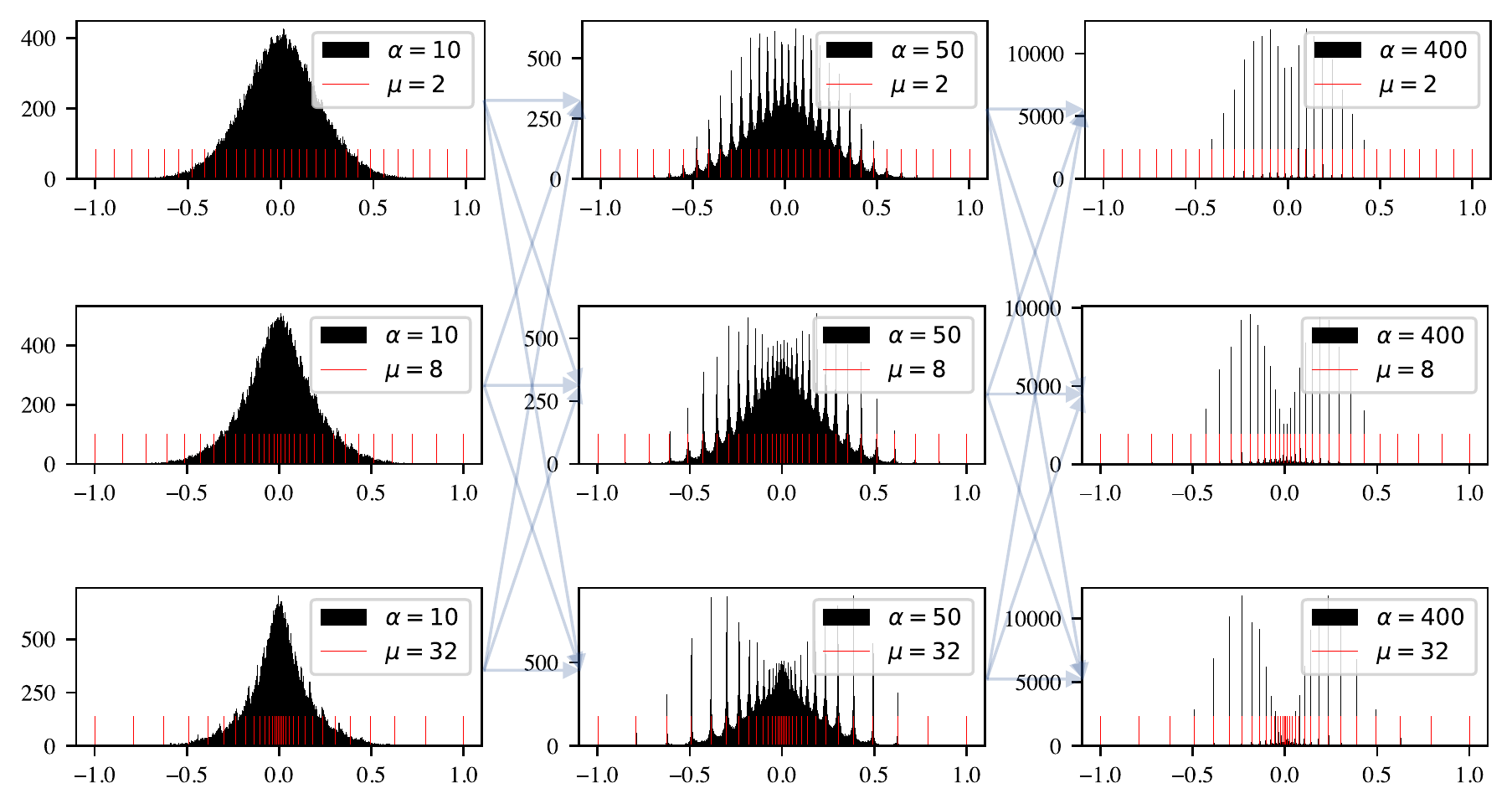}} 
\hfill
\subfigure[An illustration of a GQ callback, in which every weight matrix in an arbitrary ASR network is compressed to a certain bit-depth given the annealing factor $\alpha$ and non-linearity factor $\mu$.]{\includegraphics[width=0.42\linewidth]{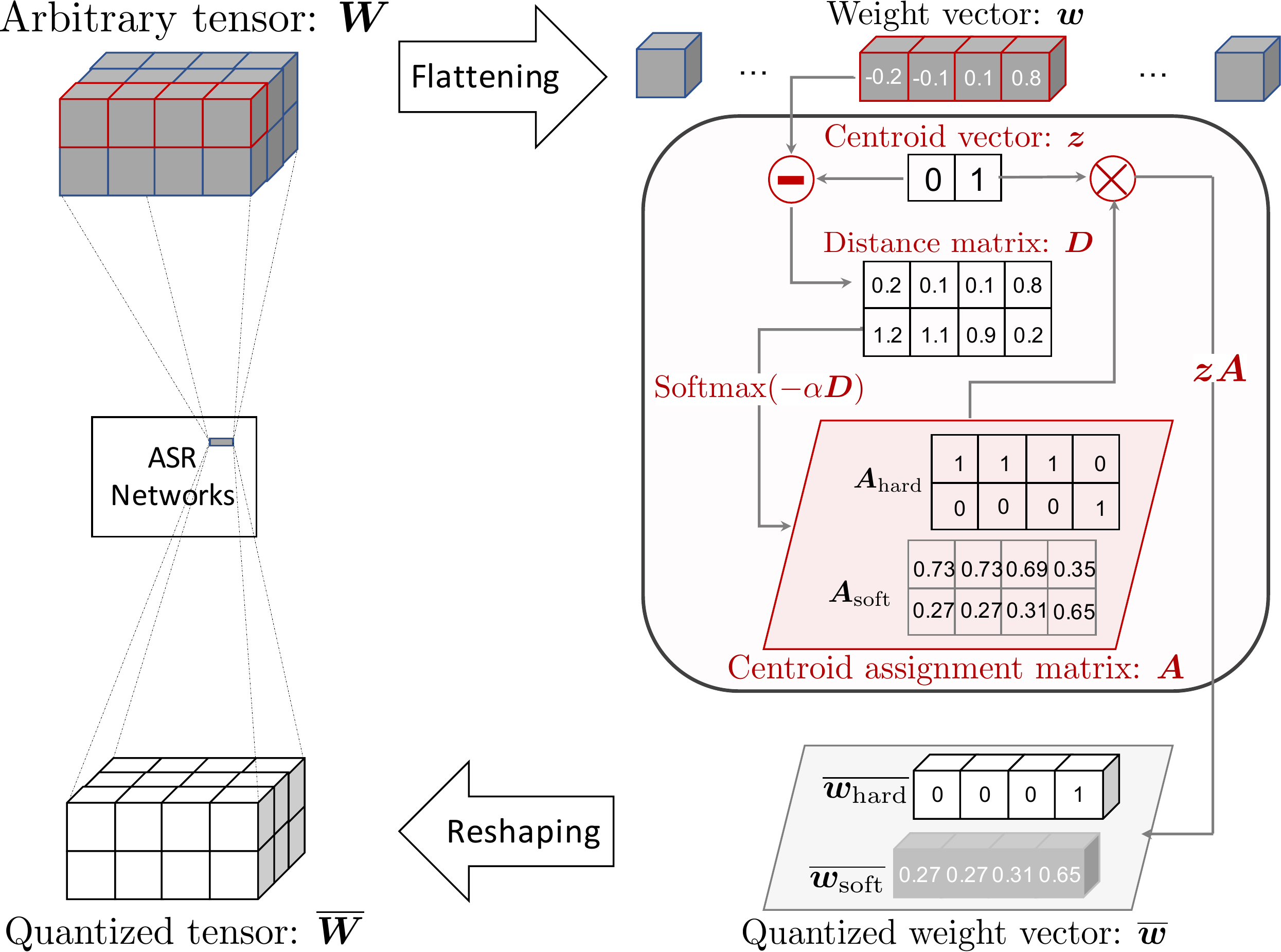}} 
	\caption{The Softmax temperature $\alpha$ and non-linearity factor $\mu$ for quantization centroids are updated during GQ callbacks.}
	\label{fig:approximation}
	\vspace{-0.12in}
\end{figure*}

\subsection{Adjusting Centroids with $\mu$-Law Expanding}

We assume weight distribution symmetry from any kernel in a trained 32-bit neural network in which the absolute values of most weights are small. Consequently, the imposed $m$ quantization centroids in $\boldsymbol{z}$ should also be symmetric ($|z_i|=|z_{m+1-i}|$ where $i\in[1,m]$) with most centroids close to 0. 
To specify and adjust the level of non-linearity of $\boldsymbol{z}$ per kernel during training, we resort to $\mu$-Law algorithm, mainly used in 8-bit PCM telecommunication (similar to $A$-Law algorithm standardized in Europe). The motivation for using $\mu$-Law function is that it accents samplings from small (soft) values, reducing the quantization error and increasing signal-to-quantization-noise-error (SQNR) for data transmission. Hence, we employ the $\mu$-Law algorithm in GQ to improve the quantization robustness of ASR models.

In $\mu$-Law expanding function (Eq. \ref{eq:inv-mu}), $\boldsymbol{z}$, linearly spaced values within the range of -1 and 1, are warped as $\boldsymbol{z'}$ in which the values are  driven closer to 0, except for the boundary poles.
 As shown in Fig. \ref{fig:s2h} (b), when $\mu$ increases, the linearly spaced values are more noticeably warped in the $\mu$-Law transformed space: 
 by adjusting the value of $\mu$ that minimizes the quantization error $\arg\min\limits_{\mu}||\boldsymbol{z'}\times{\boldsymbol{A}_{\text{soft}}}-\boldsymbol{w}||_2$, quantization centroids in $\boldsymbol{z}$ can be re-distributed to better reflect the dynamic weight range of a specific neural component. A larger $\mu$ means the weight distribution is concentrated near 0; therefore, we allocate more quantization centroids near the origin. Smaller $\mu$ values indicate the weight distribution is tail-heavy.

\begin{align}
\boldsymbol{z'} = \text{sng}(\boldsymbol{z}) \biggl( \frac{(1+\mu)^{|\boldsymbol{z}|}-1}{1+\mu} \biggr).
\label{eq:inv-mu}
\end{align}

\begin{table}[h]
\centering
\caption{The Conformer architecture with $\sim$ 28M parameters and 14 blocks. All first 13 blocks share the same topology while the last block does not include the dense kernel labeled by $^\ddagger$. Those with $^\dagger$ form the dot product multi-head attention with relative positional embeddings (causal-relmha) \cite{dai2019transformer}. Biases are not listed.}
\vspace{-0.12in}
\label{tab:topo}
\setlength\tabcolsep{6.pt}
\footnotesize
\begin{tabular}{ c|c|c|c }
 \toprule
 & Conformer Module  & Kernel Shape &Params (M) \\
 \midrule
\multirow{16}{*}{\rotatebox[origin=c]{90}{\zhenk{Encoder}}} & \begin{tabular}{@{}c@{}}Subsampling\\ Kernel\end{tabular} & 
\begin{tabular}{cc}\rule[6pt]{0pt}{0pt}(3, 3, 1, 128)  &\\(3, 3, 128, 128)  &\\ (6144, 256) &\rule[-1pt]{0pt}{0pt}\end{tabular}&1.72 \\
\cmidrule{2-4}
& \begin{tabular}{@{}c@{}}Conformer\\Block\end{tabular} &  \begin{tabular}{cc}\rule[6pt]{0pt}{0pt}\relax
(256, 1024)\hspace{10pt}\rdelim]{12}{-25mm}[$\times$14]&\\\relax
(1024, 256)&\\
(4, 256, 64)$^\dagger$ &\\
(4, 256, 64)$^\dagger$ &\\
(4, 256, 64)$^\dagger$ &\\
(4, 256, 64)$^\dagger$ &\\
(4, 64, 256)$^\dagger$ &\\
(1, 1, 256, 512)&\\
(32, 1, 256, 1)&\\
(1, 1, 256, 256)&\\
(256, 1024)&\\
(1024, 256)$^\ddagger$&\rule[-1pt]{0pt}{0pt}\end{tabular} &21.87 \\ 
\cmidrule{2-4}
& 
 \begin{tabular}{@{}c@{}}Encoder Proj\\Dense Kernel\end{tabular} &(256, 100) &0.26 \\
\midrule
\midrule
\multirow{4}{*}{\rotatebox[origin=c]{90}{\zhenk{Decoder}}}& \begin{tabular}{@{}c@{}}Single \\LSTM Layer\end{tabular} & 
\begin{tabular}{cc}\rule[6pt]{0pt}{0pt}(256, 2560)  &\\ (640, 2560)&\rule[-1pt]{0pt}{0pt}\end{tabular} &\multirow{3}{*}{{2.62}}\\
\cmidrule{2-3}
& \begin{tabular}{@{}c@{}}Vocab Proj\\Dense Kernel\end{tabular}& (640, 512)  \\
\midrule
\midrule
\multirow{1}{*}{\rotatebox[origin=c]{0}{\zhenk{Joint}}}& \begin{tabular}{@{}c@{}}Dense Kernel\end{tabular} & (512, 2501) & 1.28\\
\bottomrule
\end{tabular}
\vspace{-0.12in}
\end{table}

\subsection{
S8B Model for 8-Bit Computing}
Due to limited chip memory size and bandwidth of the NNA, weights are loaded from system memory to the chip memory per matrix, which is time consuming. Hence, compressing the model into 
S8B
can achieve inference speedup, even though NNA uses \texttt{INT8} for neural computing (see Fig.\ref{fig:nna}). 

Nonetheless, we map $\boldsymbol{z'}$ in Eq. \ref{eq:inv-mu}  to the closest value in $[\cdots k/128 \cdots]$, where the integer $k \in [-128, 127]$, such that in-training and runtime quantization centroids are consistent.


\subsection{Callback ``Is All You Need"}
A callback is a set of functions to be invoked at certain training stages.
Under GQ, the callback is all you need: For any tensor from an ASR model, during the callback, GQ will be applied to every weight vector 
$\boldsymbol{w}\in \boldsymbol{W}$ 
compressing it into 
$\overline{\boldsymbol{w}}$, for any $\boldsymbol{W} \in \mathbb{W}$.
Concretely in Fig. \ref{fig:approximation} (b), we consider a binarized case where the centroid vector contains two values: 0 and 1. As the annealing factor $\alpha$ increases during training, the centroid assignment probabilities in $\boldsymbol{A}$ become more contrastive with the probability on the closest centroid becoming almost 1 and the other nearing 0. When $\alpha$ is sufficiently large,  $\boldsymbol{A}$ approximates $\boldsymbol{A}_\text{hard}$ where each row is a one-hot vector. In the meantime, $\mu$ is adjusted as shown in Fig. \ref{fig:approximation} (a). Up to this stage, hard-quantizing $\mathbb{W}$ with $\boldsymbol{A}_\text{hard}$ does not yield noticeable degradation. In other words, without any regularizer, GQ effectively pushes weights to centroids during training for runtime model compression.

\section{Experiment}
\label{sec:exp}
\subsection{Model}
We consider both RNN-T and a lightweight streamable Conformer for experimental validation. We build a conventional RNN-T consisting of 5 LSTM encoding layers and 2 LSTM decoding layers with 1024 hidden units per layer and a fully connected joint layer. Furthermore, we benchmark GQ on an RNN-T variant with a branched encoder, named Bifocal RNN-T \cite{macoskey2021bifocal}. It has 2 encoders of different computational complexity and decides on-the-fly which encoder to use per input frame. Aside from the encoder in conventional RNN-T, bifocal RNN-T has another much smaller encoder with the same amount of layers but only 256 hidden units to process less intentful input frames.

Another end-to-end streaming ASR model is Conformer \cite{gulati2020conformer}, whose audio encoder consists of stacks of Conformer blocks (see Table.\ref{tab:topo}). Each Conformer block consists of two feedforward layers,  a multi-head attention layer, and a convolutional module. We build a Conformer with 14 layers in which we use multi-head attention with 4 heads and each head with a dimension of 64. We make it causal by applying masks to multi-head attention layers to only attend to the left context. For the sub-sampling block of Conformer, we use two layers of 2D CNN with filters of 128 channels, kernel size of 3, and stride of 2. The feedforward hidden unit dimension is 1024. 
We adopt SpecAug \cite{park2019specaugment} with the following hyper-parameters: maximum ratio of masked time frames=0.04,  adaptive multiplicity=0.04, maximum ratio of masked frequencies=0.34, and number of frequency masks=2.

We use a word-piece tokenizer and generate 2500  word-piece tokens as the output vocabulary.   
We use the Adam optimizer \cite{kingma2014adam} with $\beta_1$=0.9, $\beta_2$=0.98, and $\epsilon$=1e-9.  
The learning rate is 0.002 with 10k warm-up steps.
The step size is 1k, and 5k for Librispeech and de-identified in-house data, respectively, and the model is trained until no improvement is observed on the dev set. 
With $\mu$=8, $\alpha$ is gradually increased from 10 to 400 before the hard compression.

\subsection{Data}
We train conventional RNN-T and Conformer on LibriSpeech data corpus with 960 hours of training data for 120k steps. The 5.4 hours of dev-clean dataset and 5.3 hours of dev-other dataset are used for checkpoint selection. Models are then evaluated on 5.4 hours of test-clean and 5.1 hours of test-other dataset. We train the Bifocal RNN-T on a de-identified far-field dataset consisting of 100k hours of human transcribed data for 700k steps, validate the training via 50k utterances of dev dataset, and evaluate the model with 50k test utterances that are frequently queried in spoken language understanding tasks. We benchmark the UPL metrics from 4 NNA embedded devices on 6k utterances.

\subsection{Accuracy Comparison for RNN-T and Conformer}
For conventional RNN-T, we compare GQ with its BP-QAT \cite{nguyen2020quantization} in various
S8B
settings in Table.\ref{tab:canonical}. At 6-bit, GQ achieves a lower word error rate (WER) from all 4 datasets with 5.7\% relative WER improvement on test-clean and 2.6\% on test-other. It shows no degradation at 5-bit and 6-bit from 32-bit. GQ at 4-bit only shows less than 2\% relative 
degradation compared to QAT at 8-bit.

\begin{table}[!t]
	\centering
	\footnotesize
	\caption{WER performance of Conventional RNN-T on LibriSpeech datasets}\vspace{-0.12in}
	\setlength\tabcolsep{7.pt}

		\begin{tabular}{ l | c | c|c|c}
			\toprule
			& dev-clean  & dev-other& test-clean & test-other
			\\
			\midrule
			32-bit baseline & 8.11 & 21.27 & 8.68 & 22.29 \\\midrule
            8-bit QAT & 8.15 & 21.41 & 8.70 & 22.36 \\\midrule
			6-bit QAT & 8.32&  21.84&  8.90 & 22.82 \\\midrule
			\midrule
			6-bit GQ & 7.76 & 20.80 & 8.39 & 22.12 \\\midrule
            5-bit GQ & 7.93 & 21.34 & 8.33 & 22.16 \\\midrule
            4-bit GQ & 8.23 & 21.82 & 8.78 & 22.54\\
			\bottomrule
		\end{tabular}
	\label{tab:canonical}
	\vspace{-0.12in}
\end{table}

We also report WERs from a GQ compressed Conformer in Table. \ref{tab:conformer_lib}. While the 32-bit baseline yields the best WER on the training set, both 5-bit and 6-bit quantized Conformers (with all other settings being the same) generalize better on dev-clean and dev-other datasets.
Furthermore, we select the best checkpoint based on dev-clean WER and observe no degradation from the test sets with the model size reduced by 6.4$\times$.   Although this seems counter-intuitive in that the compressed model outperforms the 32-bit baseline, it is not rare as also shown in \cite{ding20224}. One explanation is that by driving weights towards quantization centroids, the search space is drastically reduced, yielding an arguably easier optimization process.

Although GQ shows little ($<$ 5\% relatively) to no accuracy loss compared to the 32-bit baseline, it is worth mentioning that 4-bit quantization severely impacts the generalizability of Conformer on dev and test datasets, compared to the 5-bit mode. Even when only multi-head self-attention (MHSA) modules (approximately only 15\% of total parameters) are quantized in 4-bit, the WER from dev and test datasets declines noticeably while that from the training dataset is much better. It is also observed from the 4-bit conv-block setting where about 77\% of weights are 4-bit compressed.

\begin{table}[!t]
	\centering
	\footnotesize
	\caption{GQ performance on a lightweight and streaming Conformer for LibriSpeech datasets. Here, full x-bit means all weights in the model are compressed to x-bit. If a module is labeled as x-bit, all weights in that module are compressed to x-bit with all other weights in 8-bit.}\vspace{-0.12in}
	\setlength\tabcolsep{1.pt}
		\begin{tabular}{ l |c| c | c|c|c|c}
			\toprule
			& train& dev-clean  & dev-other& test-clean & test-other & size reduct.
			\\
			\midrule
			32-bit baseline &0.86& 5.62 & 14.11 & 5.74 & 14.21&-- \\\midrule
			\midrule
			Full 6-bit&1.42& 5.06 & 13.42 & 5.26 & 13.38&5.3$\times$ \\\midrule
            Full 5-bit&2.15& 5.38 & 13.58 & 5.50 & 13.85&6.4$\times$\\\midrule
            4-bit Conv-Block & 1.73 & 5.97 & 14.52 & 6.02 &14.87&7.1$\times$\\\midrule
            4-bit MHSA & 1.02 & 5.46 & 13.85 & 5.76 & 14.26&4.6$\times$\\
			\bottomrule
		\end{tabular}
	\label{tab:conformer_lib}
	\vspace{-0.12in}
\end{table}

\subsection{Accuracy, Memory Footprint and UPL Comparisons for Bifocal RNN-T}
To better understand GQ's impact on memory footprint and UPL, 
we apply both GQ and our previous QAT methods to Bifocal RNN-T trained on a de-identified far-field dataset, where weights are compressed to various bit-depth configurations (see Table.\ref{tab:comp}). Under S8B-QAT, we compress weights in all but the first layer in 5-bit for the left encoder (L-Enc), right encoder (R-Enc), and decoder with all other weights in 8-bit. With GQ, we compress the model to 5-bit (5B-GQ) or lower (S5B-GQ) without damaging the predictive performance on the frequent test set, compared to our previous 8-bit (8B-QAT) and sub-8-bit (S8B-QAT) methods. 

We compile the ONNX files \cite{bai2019} of the trained models to hardware executable binary files for memory and UPL benchmarkings. In Table. \ref{tab:comp}, the memory consumption is reduced to 20.83MB in 5-bit from 30.07MB in 8-bit, which amounts to a $30.73\%$ memory savings and yields to 20\% p50 UPL reduction (32.30\% and 32.75\% UPL reduction for P90 and P99, respectively). Although we binarize the left encoder without degrading the accuracy, the impact on memory and latency is not significant as the left encoder is already small.

\begin{table*}[h]
\normalsize
	\centering
	\captionsetup{width=\linewidth}
	\caption{Memory footprint, UPL and accuracy benchmarks for Bifocal RNN-T under various quantization bit-depth settings. The number of model parameters is the same among four experimental settings for a fair comparison.}\vspace{-0.12in}
	\footnotesize
	\setlength\tabcolsep{2.5pt}
\begin{tabular}{c|cccc|cc|cccccc|c}
\toprule
       & \multicolumn{4}{c|}{Bit-Depth}        & \multicolumn{2}{c|}{Normalized Memory Footprint}          & \multicolumn{6}{c|}{Normalized UPL} & Normalized Accuracy  \\
       \midrule
& L-Enc & R-Enc & Dec & Joint  & Total &  Rel. Dgrd.& P50 &   Rel. Dgrd.      & P90 &   Rel. Dgrd. & P99     &   Rel. Dgrd.        & Frequent Test Set
\\\midrule
8B-QAT  & 8 & 8 &8 &8                  & 1.00 &--& 1.00           &--  & 1.55      &--        & 2.58       &--      & 1.00                 \\\midrule
S8B-QAT & 5/8&5/8&5/8  & 8       & 0.83 &-17.43& 0.87   &-12.52           & 1.22      &-21.08        & 2.04    & -20.68         & 0.99                 \\\midrule
5B-GQ   & 5 & 5 &5 &5                 & 0.69 &-30.73 & 0.80   &-19.75           & 1.07       &-31.30       & 1.76      &-31.75       & 0.96                 \\\midrule
S5B-GQ  & 1        & 5&5&5      & 0.68 &-32.09& 0.80   &-20.31          & 1.05      &-32.20       & 1.70       &-34.07      & 0.97            \\      \bottomrule
\end{tabular}
\label{tab:comp}
\vspace{-0.12in}
\end{table*}

\begin{figure*}
     \centering
\subfigure[Quantization frequency effect on training WER. The effect on other datasets is similar.]{\includegraphics[width=0.3\linewidth]{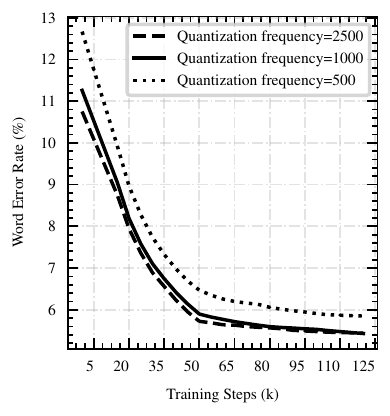}} 
\hfill
\subfigure[Kernel-wise bit allocation and its weight boundary for the 6-bit Conformer. Every quantized kernel contains no more than 64 distinct weight values. Some kernels are with a narrow weight boundary and consume less than 32 values, indicating a possibility of 5-bit quantization.]{\includegraphics[width=0.65\linewidth]{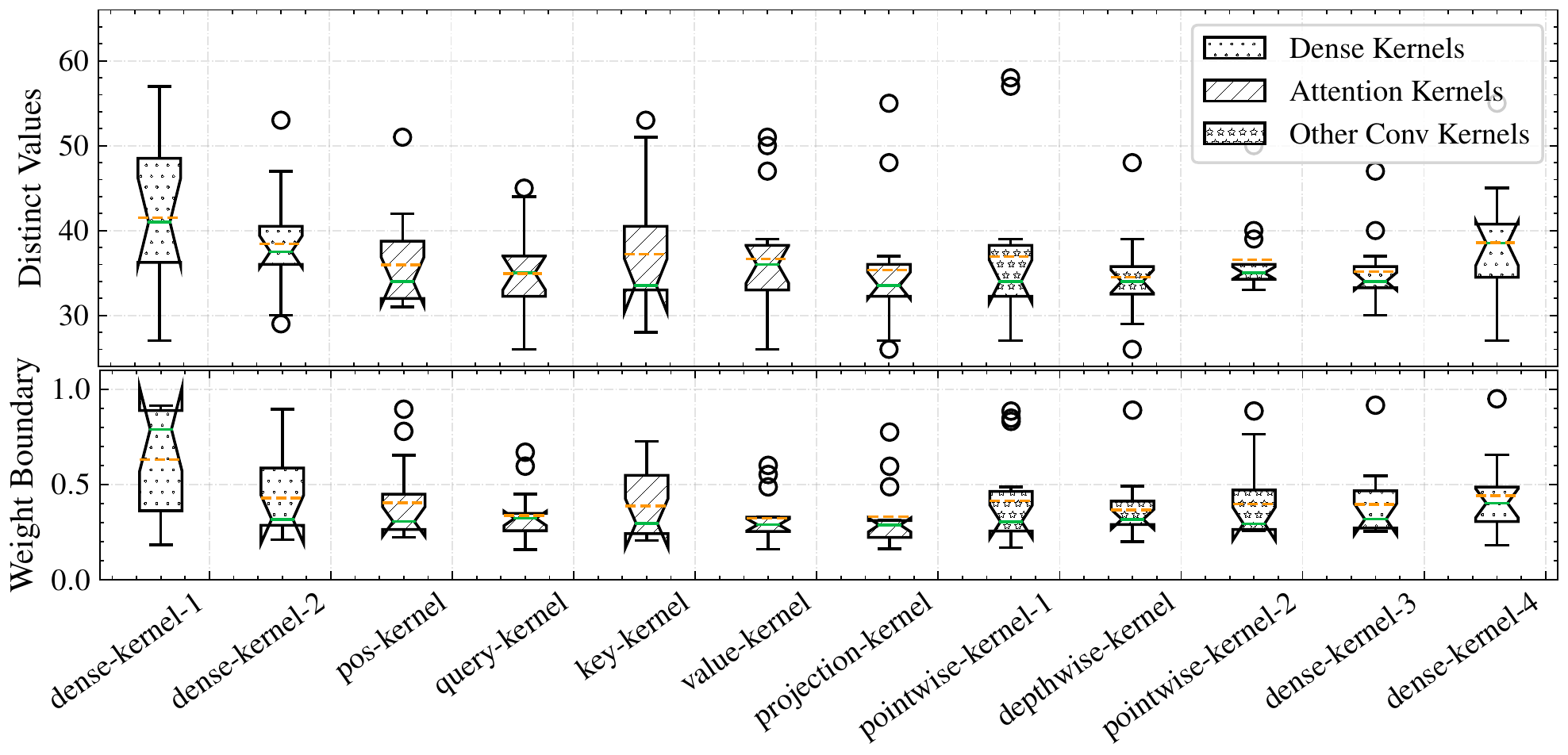}} 
\vspace{-0.12in}
        \caption{Without loss of generality, we use Conformer for ablation studies on quantization frequency and bit allocation of GQ.}
        \label{fig:freq}
        \vspace{-0.12in}
\end{figure*}

\subsection{Analysis of Quantization Frequency and Bit-Allocation}
To understand how frequent should GQ be invoked during training, we alternate different quantization frequencies to train the 6-bit Conformer. Fig. \ref{fig:freq} (a) shows that a too frequent in-training quantization setting hurts the predictive performance. One explanation is that it offsets the gradient effect by dragging weights back to the near-centroid zone. In practice, we observe no accuracy degradation even when the quantization frequency is 10k training steps, allowing us to compress a Conformer model to 4-bit in just a few GQ invocations. Additionally, the WER curves in Fig. \ref{fig:freq} (a) are relatively smooth as GQ periodically performs, which indicates that the training is quantization aware even without any augmented regularizers.

We perform bitwise analysis of all encoding blocks of a lightweight and streaming Conformer (Fig. \ref{fig:freq} (b)). 14 encoding blocks are included in the Conformer, accounting for $\sim80\%$ of the total parameters. The Conformer is fully quantized to 6-bit requiring that all weight matrices (kernels) must have no more than 64 distinct weight values. Weight matrices are categorized as  a dense kernel, MHSA kernel, or other types of convolution kernel, such as a depthwise convolution kernel.
Using a box plot, we show the lower and upper quartiles of the 95\% confidence interval along with the mean (orange dotted line) and median (green solid line) values for weight matrices from all blocks. It is worth noting that most weight matrices do not consume all 64 distinct values from 6-bit quantization to yield predictive results as good as 32-bit counterparts. Particularly, convolution kernels, whether in MHSA modules or not, are less bit consuming than dense kernels. 
The number of distinct values is found positively correlated to the weight boundary of the kernel: a smaller weight range indicates a fewer number of distinct values or quantization bit-depth and vice versa. 
To further reduce the memory footprint and UPL, 
compressing components with a narrow boundary to a lower bit-depth could be preferred over the dense kernel with a larger weight boundary.
\section{Concluding Remarks}
\label{sec:conclude}
We proposed General Quantizer (GQ) a plug-and-play QAT mechanism, allowing models to be compressed to an arbitrary bit-depth during training without augmented regularizers. We applied GQ to three popular end-to-end Automatic Speech Recognition (ASR)
 models: conventional RNN-T, Bifocal RNN-T, and Conformer. In various sub-8-bit settings, GQ shows little to no accuracy degradation while noticeably reducing the memory footprint and user-perceived latency. GQ is model-agnostic and can be applied to 
 feature map compression as one future direction.

%

\bibliographystyle{IEEEbib}
\bibliography{refs19}

\end{document}